\documentclass[12pt]{iopart}
\usepackage{graphicx}
\begin{document}
\bibliographystyle{unsrt}
\setlength{\baselineskip}{0.5cm}
\newcommand{\be}{\begin{eqnarray}}
\newcommand{\ee}{\end{eqnarray}}
\newcommand{\xx}{\begin{eqnarray*}}
\newcommand{\yy}{\end{eqnarray*}}
\newcommand{\nn}{\nonumber}         
\newcommand{\m}{{\bf M}} 
\newcommand{\al}{\alpha} 
\newcommand{\bt}{\beta} 
\newcommand{\lb}{\lambda} 
\newcommand{\vp}{{\vec \phi}} 
\newcommand{\Real}{{\rm Re}} 
\newcommand{\Imag}{{\rm Im}} 
\newcommand{\rf}[1]{$(\ref{#1})$} 
\newcommand{\rfb}[1]{$\left [{\rm equation}\;(\ref{#1})\right ]$} 
\newcommand{\figref}[1]{{\rm figure\;$\ref{#1}$}}

\title{Asymptotic behavior of the density of states on a random lattice}
\author{Jean-Yves Fortin}
\address{CNRS, Laboratoire de Physique Th\'eorique, UMR7085,
3 rue de l'Universit\'e, 67084 Strasbourg Cedex, France}
\date{\today}
\ead{fortin@lpt1.u-strasbg.fr}
\begin{abstract}
 We study the diffusion of a particle on a random lattice with
fluctuating local connectivity of average value $q$. 
This model is a basic description of relaxation processes in 
random media with geometrical defects.
We analyze here the asymptotic behavior of the eigenvalue distribution 
for the Laplacian operator. We found that the localized states outside 
the mobility band and observed by Biroli and Monasson, 
in a previous numerical analysis~\cite{biroli.99}, 
are described by saddle point solutions that
breaks the rotational symmetry of the main action in the real space. 
The density of states is characterized asymptotically by a series of 
peaks with periodicity $1/q$.

\end{abstract}
\maketitle

Diffusion on random graphs can be a useful problem for studying relaxation 
processes in glassy systems in general. Usually, the disorder arises 
from a random potential, impurities, but a random geometry can also play this 
role~\cite{campbell.85}. One can visualize a diffusion
of a particle on a random graph as the relaxation of a disordered system
out of equilibrium on a complicate energy landscape.
For example, this relaxation in random Ising magnets can be 
identified with diffusion on the vertices of a hypercube in 
the configuration space, and the edges correspond to the energy
paths that connect one configuration to another~\cite{ogielski.85}. 
 Numerical simulations~\cite{ogielski.85,ogielski.87} in Ising spin glasses
of the order parameter $q(t)=[<S_i(t)S_i(0)>]_{av}$, where $<\ldots>$
is the thermal average and $[\ldots ]_{av}$ the average over disorder,
show that this quantity follows a Kohlrausch law similar to a ``stretched'' 
exponential $\exp(-(t/\tau)^{\beta})$
with $1/3\le \beta\le 1$ in the region just above the spin glass phase.
This kind of non-exponential relaxation is typical of many glassy systems
\cite{campbell.85,campbell.88,svedlindh.87},
unlike the usual exponential behavior with only one relaxation time.
 The coefficient $\beta$ varies with temperature from 1 in the high 
temperature phase, to $1/3$ at the glassy transition. In between, 
there seems to be a phase of localized states, or Griffiths phase, where
$\beta$ is in between from 1 to 1/3. The value 1/3 seems to be universal in 
many experimental systems and has been observed in spin glass 
${\rm Eu_{0.4}Sr_{0.6}S}~$\cite{bontemps.88}, at the glass transition 
of the crystalline fast ion conductor Na $\beta$-alumina~\cite{ngai.83}, 
or molten salts ${\rm CaK(NO_3)}$~\cite{ngai.84}. 
The question about the universality of the Kohlrausch law has
been raised, and exact dynamics on ultra-metric spaces~\cite{ogielski.prl.85}
show that it depends on the scaling of the barriers with the distance
between different states. It could be algebraic (linear dependence 
with the distance) or Kohlrausch like (logarithm dependence).
 Random graph models can also be a useful tool to study propagation of 
sound waves in granular and random media~\cite{liu.92,jia.99}, since diffusion,
quasi-diffusion and localization of sound waves are similar to diffusion 
on a random network, the edges being the connections between the 
particles making up the medium. Frequency response~\cite{jia.99} for 
different amplitudes shows a region of coherent propagation plus a 
quasi-diffusive regime. Numerical and simplified models~\cite{leibig.94} 
of a granular medium for small vibration amplitudes show modes that are 
extended and localized in space, like a particle moving on a random medium.

 In order to simulate the glassy systems discussed above with short-range 
interactions and also percolation problems, Viana, Rodgers and Bray (VRB) 
studied a simple model of diffusion based on sparse 
random matrix~\cite{viana.85,bray.88}. Their model has the property to have 
long-range interactions with dilute coordination numbers,
so that in average the coordination number is finite and we may expect
to have a closer view of some experimental material. 
The spectrum of eigenvalues (all positive) spreads over a continuous
band, and for the particular case of the Cayley tree model, are bounded 
between mobility edges, with a gap from below (see also \cite{kim.85}). 
 Recently, Biroli and Monasson (BM)~\cite{biroli.99} studied the same model 
numerically, and showed that localization effects arise outside the 
extended region, with peaks at some regular intervals where the inverse
partition number is high, showing that these peaks are resonance for
localization. The existence
of a Griffiths region was analyzed by Rodgers and Bray~\cite{rodgers.88}
who found a tail distribution for large eigenvalues outside the mobility
band (with a random matrix containing only elements -1, 0 and 1)
but there is no peak structure in their analysis contrary to BM's numerical
result. Their saddle point solution is invariant by rotation in both 
replica and real spaces and they do not discuss the possibility 
of the rotational symmetry breaking in at least one of these spaces. 
We will analyze in this paper the possibility of a real space symmetry 
breaking for the site dependent fields, unlike the method of self-consistent
field equations developped in \cite{rodgers.88}. 
It is clear that in order to describe 
the localized regime, we have to take account of this kind of solutions.
 Our motivation is therefore to find the asymptotic solutions in 
the region of the density of states outside the mobility 
band for the VRB's model.  
The original model of relaxation in a geometrical disordered system consists of
$N$ points connected randomly by bonds $M_{ij}=-1/q$ with a probability
equal to $q/N$ and $M_{ij}=0$ else. 
This is an infinite range model, but the average
coordination number $q$ is finite and we may expect that the 
dilute exchange interactions make the model similar to a short range model
\cite{kanter.87}. We follow the 
references~\cite{biroli.99} and~
\cite{bray.88} for the notations.
On a random lattice, a particle performs a random walk from one point to
another.  
Let $c_i(t)$ be the probability for the particle to be on the site $i$ 
at time $t$.
Then the master equations describing the time evolution of these 
amplitudes are

\be\label{eq1}
\frac{d c_i}{dt}=-\sum_{j}M_{ij}c_j
\ee

where the probability for the symmetric elements $M_{ij}$ is

\be\nn
P(M_{ij})=\frac{q}{N}\delta (M_{ij}+\frac{1}{q})+
(1-\frac{q}{N})\delta (M_{ij}),
\ee

for $i\neq j$. The diagonal elements are equal to $M_{ii}=-\sum_{j\neq i}
M_{ij}$. This ensures that the eigenvalues of the matrix $M$ are all 
positive since the quadratic form 

\be\nn
\sum_{i,j}M_{ij}x_ix_j=-1/2\sum_{i,j}M_{ij}(x_i-x_j)^2
\ee

is always positive~\cite{bray.88}. 
Let $\lambda_i$ be an eigenvalue of $\m$, with ${\bf V}_i$ the corresponding 
eigenvector, then the general solution of equation \rf{eq1} is 

\be
c_i(t)=\sum_{j,k}({\bf V}_j)_i({\bf V}_j)_kc_k(0)\exp(-\lambda_j t).
\ee

If the particle is on the site $m$ at $t=0$, i.e.
$c_i(0)=\delta_{i,m}$, then the probability that after a time $t$ the particle
is found on the same site is equal to $c_m(t)$, and we can define
an average probability of return to the origin at time $t$ after 
summing up over all the possible origin points:

\be\nn
f(t)&=&\left [\frac{1}{N}\sum_{m}c_m(t)\right ]_{av}=
\left [\frac{1}{N}\sum_{m,j}({\bf V}_j)_m({\bf V}_j)_m
\exp(-\lambda_j t)\right ]_{av}
\\ \label{eq18}
&=&\left [\frac{1}{N}\sum_{j}\exp(-\lambda_j t)\right ]_{av}
=\int_0^{\infty} P(\lambda)\exp(-\lambda t)d\lambda,
\ee

where $[\ldots ]_{av}$ is the average over the link configurations. 
$P(\lb)$ is the probability distribution function for the eigenvalues
of the symmetrical matrix $\m$. In figure 1 we have solved
numerically $P$ by diagonalizing 4500 different samples of random matrices 
$\m$ with the parameters $N=800$ and $q=20$ 
(see also~\cite{biroli.99}). Then we made an histogram of all the 
different eigenvalues found and normalized the distribution.
Following the results from the partition number of reference \cite{biroli.99}, 
the distribution presents a mobility band between approximatively $\lb=0.5$
and $\lb=1.7$ plus localization edges outside this band and containing peaks.
The distribution $P(\lambda)$ can be expressed as a functional of 
replicated fields $\phi_i^{\al}$. Indeed, we can write

\be\nn
P(\lambda)&=&\left [ \frac{1}{N}\sum_i\delta(\lb-\lb_i)\right ]_{av}
\\ \nn
&=&\left [\frac{1}{N}\tr \delta(\lb-\m)\right ]_{av}=
\left [-\Imag\frac{1}{\pi N}\tr G(\lb +i\epsilon)\right ]_{av},
\ee

where we introduce the Green function $G(\lb+i\epsilon)=1/(\lb+i\epsilon-\m)$. 
More precisely:

\be \nn
P(\lb)&=&\left [-\Imag\frac{1}{\pi N}\frac{\partial}{\partial \lb}
\tr \log (\lb+i\epsilon-\m)\right ]_{av}
\\ \nn
&=&\left [ -\Imag\frac{1}{\pi N}\frac{\partial}{\partial \lb}
\log \det (\lb+i\epsilon-\m) \right ]_{av}.
\ee

The determinant can be rewritten as a Gaussian integral over 
fields $\phi_i$, $i=1,\ldots,N$:

\be\nn
P(\lb)&=&\left [\Imag\frac{2}{\pi N}\frac{\partial}{\partial \lb}
\log \int \prod_{i=1}^{N}d\phi_i\exp\left (
\frac{i}{2}\sum_{i}(\lb+i\epsilon)\phi_i^2 \right )\right .
\\
& &\left .\exp \left (-\frac{i}{2}\sum_{i,j}\phi_iM_{ij}\phi_j \right ) 
\right ]_{av}.
\ee

 The average over the link configurations is performed by using 
replicated fields $\phi_i^{\alpha}$, and we obtain (see \cite{bray.88}
for the details):

\be\nn
P(\lb)&=& \left [\Imag\frac{2}{\pi N}\frac{\partial}{\partial \lb}
\log Z \right ]_{av}=\lim_{n\rightarrow 0}
\Imag\frac{2}{\pi nN}\frac{\partial}{\partial \lb}
\left [ Z^n \right]_{av}
\\ \nn
&=&\lim_{n\rightarrow 0}
\Imag\frac{2}{\pi nN}\frac{\partial}{\partial \lb}
\int \prod_{i,\alpha}d\phi_i^{\alpha}\exp \left (\frac{i}{2}\sum_{i,\alpha}
(\lb+i\epsilon)\phi_i^{\alpha 2} \right )
\\ \nn
& &\exp \left (
\frac{1}{2}\sum_{i,j}\log
\left [ 1-\frac{q}{N}+\frac{q}{N}\exp -\frac{i}{2q}\sum_{\alpha}
(\phi_i^{\alpha}-\phi_j^{\alpha})^2 \right ] \right )
\\ \label{eq2}
&=&
\lim_{n\rightarrow 0}
\Imag\frac{2}{\pi nN}\frac{\partial}{\partial \lb}
\int \prod_{i,\alpha}d\phi_i^{\alpha}\exp S(\lb,\{\phi_i^{\al}\}).
\ee

Now we want to study the asymptotic behavior of the distribution $P$
by evaluating the saddle points of $S$ in the limit $\lb\to\infty$,
while we keep $N$ large but fixed first. It seems very similar to studying the 
saddle point solutions of the extensive function $S$ in the large $N$ limit,
but the difference is that we are studying the asymptotics of the distribution
which does not describe the mobility part of the curve, so the 
corresponding saddle points are different. 
 Moreover we are looking for solutions which can break the 
rotational space symmetry, since we are interested in the localized 
regime where strong or low connectivity may contribute to the distribution. 
A particle localized in one region would favor the amplitudes of the sites
inside this region.
The extrema of the functional $S$ in equation \rf{eq2} are given by 
the set of equations 

\be\label{eq3}
(\lb+i\epsilon)\phi_i^{\al}=\frac{1}{N}\sum_{j}
\frac{
\left (
\phi_i^{\al}-\phi_j^{\al}\right )
\exp\left (-\frac{i}{2q}\sum_{\beta}(\phi_i^{\beta}-\phi_j^{\beta})^2
\right )
}
{
1-q/N+q/N\exp\left (-\frac{i}{2q}\sum_{\beta}(\phi_i^{\beta}-\phi_j^{\beta})^2
\right )}.
\ee

The simplest solution is when all fields have the same value
on every site $\phi_i^{\al}=\phi^{\alpha}$, which directly leads to 
$\phi_i^{\al}=\phi^{\al}=0$. This gives a Dirac function centered at 
$\lb=1$ and this is not an asymptotic solution for our problem.
The next step is to take $\phi_i^{\al}=\phi^{\alpha}$ everywhere 
except on one site $i_0$ where $\phi_{i_0}^{\al}\ne \phi^{\alpha}$.
We obtain 2 equations (up to the order $1/N$):

\be\nn
(\lb+i\epsilon)\phi_{i_0}^{\al}=\frac{N-1}{N}
\left ( \phi_{i_0}^{\al}-\phi^{\al}\right )
\exp\left (-\frac{i}{2q}\sum_{\beta}(\phi_{i_0}^{\beta}-\phi^{\beta})^2
\right ),
\\ \label{eq3bis}
(\lb+i\epsilon)\phi^{\al}=\frac{1}{N}
\left ( \phi^{\al}-\phi_{i_0}^{\al}\right )
\exp\left (-\frac{i}{2q}\sum_{\beta}(\phi^{\beta}-\phi_{i_0}^{\beta})^2
\right ).
\ee

In the large $N$ limit, only the field $\phi_{i_0}^{\al}$ does not vanish
($\phi^{\al}=-\phi_{i_0}^{\al}/(N-1)$), with the length of the vector 
$(\phi_{i_0}^1,\ldots,\phi_{i_0}^n)$ satisfying the equation:

\be\label{eq4}
\phi_m^2=
\sum_{\al}\phi_{i_0}^{\al 2}=2iq\log\left (\lb+i\epsilon \right )+4\pi qm,\;
{\rm and}\;
\phi_{i\neq i_0}^{\beta}\simeq-\frac{\phi_{i_0}^{\beta}}{N},
\ee

where $m$ is an integer. There is therefore an infinite set of non
zero solutions in the complex plane. In the following we choose the 
site located at $i_0=1$ for simplification, since the solution is invariant
for the $N-1$ other sites. 
For each integer $m$, the number of all possible vectors with module 
$\phi_m^2$ is equal to $N2\pi^n/\Gamma(n/2)\sim Nn$ in the limit $n
\to 0$. Given the solution \rf{eq4}, the value of the action is equal to 

\be\label{eq5}
S_m=-q(\lb+i\epsilon)\log(\lb+i\epsilon)+q(\lb+i\epsilon)-q+2i\pi q
(\lb+i\epsilon)m
\ee

then $P$ falls exponentially with $\lb\log (\lb)$, and the corrections to 
the exponential are given by computing the fluctuations around the 
saddle points.
Unfortunately the matrix of the second derivatives of $S$ at the site $i_0=1$ 
has many zero eigenvalues and we need to consider the next order.
Indeed, we have

\be\nn
\frac{\partial^2 S}{\partial \phi_1^{\al}\partial\phi_1^{\beta}}
&=&-\frac{\lb+i\epsilon}{q}\phi_1^{\al}\phi_1^{\beta},
\;\;\frac{\partial^2 S}{\partial \phi_i^{\al}\partial\phi_j^{\beta}}
=\frac{i}{N}\delta_{\al \beta},\;\;i\neq j,
\\ \label{eq6}
\frac{\partial^2 S}{\partial \phi_i^{\al}\partial\phi_i^{\beta}}
&=&i(\lb+i\epsilon-1)\delta_{\al \beta},\;\;i,j\neq 1,
\ee

The first matrix in \rf{eq6} has $(n-1)$ zero eigenvalues, and we therefore 
have to take account of the third derivatives at the site $i_0$:

\be\nn
\frac{\partial^3 S}{\partial \phi_1^{\al}\partial\phi_1^{\beta}
\partial\phi_1^{\gamma}}
=&-&\frac{\lb+i\epsilon}{q}(\phi_1^{\al}\delta_{\beta\gamma}+
\phi_1^{\beta}\delta_{\al\gamma}+\phi_1^{\gamma}\delta_{\al\beta})
\\ \label{eq7}
&+&\frac{i(\lb+i\epsilon)}{q^2}\phi_1^{\al}\phi_1^{\beta}\phi_1^{\gamma}.
\ee

The fluctuations around the saddle point solutions $\phi_m$ 
lead to the following integrals for main contribution at large $N$:

\be\nn
I_m(\lb)=\int \prod_{i,\al}d x_i^{\alpha}
\exp \left ( -\frac{\lb+i\epsilon}{2q}\phi_1^{\al}\phi_1^{\beta}x_1^{\al}
x_1^{\beta}+\frac{i}{2}(\lb+i\epsilon-1)\sum_{i\neq 1}x_i^{\al}x_i^{\al}
\right .
\\ \nn
\left .
+\frac{i}{2N}\sum_{i\neq j}x_i^{\al}x_j^{\al}-
\frac{\lb+i\epsilon}{2q}\phi_1^{\al}x_1^{\al}x_1^{\beta}x_1^{\beta}
+\frac{i(\lb+i\epsilon)}{6q^2}\phi_1^{\al}\phi_1^{\beta}\phi_1^{\gamma}
x_1^{\alpha}x_1^{\beta}x_1^{\gamma}+\ldots
\right ).
\ee

 The Gaussian integration over the variables $x_i^{\al}$ gives 
$(2\pi/i(\lb+i\epsilon))^{n(N-1)/2}$. In the following we will set 
$\epsilon=0$, since it is not essential 
for the remaining calculation. It is useful to work in the basis where 
the quadratic terms are diagonal. 
We first set $\phi^{\al}=\phi_m\varphi^{\al}$ where 
$(\varphi^{1},\ldots,\varphi^{n})$ is a real unit vector, and $\phi_m$ a square
root of \eref{eq3}. 
The matrix $\varphi_{\al}\varphi_{\beta}$ has one eigenvalue unity
with eigenvector ${\bf V}^1=(\varphi^{1},\ldots,\varphi^{n})$ and $n-1$ 
eigenvalues zero with eigenvectors ${\bf V}^{\beta},\beta=2,..n$ 
satisfying $\sum_{\al}\varphi_{\al}V^{\beta}_{\al}=0$. We then use new
variables $y_{\al}=\sum_{\beta}V^{\alpha}_{\beta}x_1^{\beta}$ that 
diagonalize the quadratic terms. The matrix 
$V^{\beta}_{\al}$ satisfies $^t{\bf V}={\bf V}^{-1}$, so
that $x_1^{\al}=\sum_{\beta}V^{\beta}_{\al}y_{\beta}$. After some algebra, 
$I_m$ can be finally written as

\be\nn
I_m(\lb)&=&\left (\frac{2\pi}{i\lb}\right )^{n(N-1)/2}
\int \prod_{\al}d y_{\alpha}
\exp \left ( -\frac{\lb}{2q}\phi_m^2y_1^{2}
\right .
\\ & &\left .
-\frac{\lb}{2q}\phi_m y_1\sum_{\al}y_{\al}^2
+\frac{i \lb}{6q^2}\phi_m^3 y_1^{3}+\ldots
\right ).
\ee

When $n$ is close to zero the first coefficient on the right hand
side is unity. 
We then formally integrate over $y_{\al}$ for $\al\neq 1$, so that we 
are left with only one integral in the limit $n\rightarrow 0$:

\be\label{eq9}
I_m(\lb)&=&\int d y_{1}\sqrt{\frac{\lb\phi_m y_1}{2\pi q}}
\\ \nn & &
\exp \left ( -\frac{\lb}{2q}\phi_m^2y_1^{2}
+\frac{\lb}{2q}\phi_m\left (\frac{i}{3q}\phi_m^2-1\right ) y_1^{3}+\ldots
\right )
\ee

To compute this integral, we can still try to apply a saddle point method to
the function inside the exponential. This function is proportional
to $\lb$ and the first derivative vanishes for two solutions, $y_a=0$ 
and $y_b=2q\phi_m/(i\phi_m^2-3q)$. The values of the second derivatives
are equal respectively to minus and plus $\lambda\phi_m^2/q$.
If we expand the integral around $y_a$ for example, we obtain

\be\label{eq10}
I_m(\lb)=\frac{(1+i)}{\sqrt{\pi}}
\left (\frac{2q}{\lb}\right )^{1/4}\frac{1}{\phi_m}
\int_0^{\infty}dz\sqrt{z}\exp(-z^2)+\ldots
\ee

with $\int_0^{\infty}dz\sqrt{z}\exp(-z^2)=\Gamma(3/4)/2\simeq 0.612\;708$.
The density of states is finally asymptotically equal to 

\be\label{eq11}
P(\lb)&=&\Imag \frac{2}{\pi}\frac{\partial}{\partial \lb}
\left \{\exp\left (
-q\lb\log \lb+q\lb-q
\right )\right .
\\ \nn
& &
\left .\sum_{m=-\infty}^{+\infty}I_m(\lb)\exp\left ( 2i\pi q\lb m \right )
\right \}
\ee

The approximation \rf{eq10} is not accurate because we have considered only 
one saddle point, but it basically shows that for large $m$, $I_m$ decreases
like $1/\sqrt{|m|}$, so that the series in \eref{eq11} is convergent, except
for the points where $q\lambda$ is an integer. At these points the series
diverges, and we obtain peaks in the distribution. 
A more precise computation of \rf{eq9} is to find a path for which
the integral is convergent. A first transformation is to make the cubic
term inside the exponential purely imaginary, so that the integral is
defined on the real axis. This is done by setting 
$z=y_1/a$ such that the coefficient $a$ satisfies for example

\be\label{eq12}
a^3(-\frac{\lb}{2q}\phi_m+\frac{i \lb}{6q^2}\phi_m^3)=i
\ee

The integration path is then along the direction given by the vector $a$. 
There are 3 solutions for \eref{eq12}, each being proportional 
to $\exp(2i\pi k/3)$, with $k=0,1,2$. We will note them $a_{k,m}$. 
Now the coefficients of the quadratic term in \rf{eq9} is equal to 
$b_{k,m}\equiv \lb\phi_m^2a_{k,m}^2/2q$. Therefore $I_m$ can expressed as

\be
I_m(\lb)=\sqrt{\frac{\lb\phi_m}{2\pi q}}a_{k,m}^{3/2}
\exp(i\sigma \pi/4)\psi_{\sigma}(b_{k,m})
\ee

with the function

\be\label{eq13}
\psi_{\sigma}(b_{k,m})
=\sqrt{2}\int_0^{\infty}dz\sqrt{z}\exp(-b_{k,m}z^2)\left (\cos z^3 
+\sigma\sin z^3 \right ).
\ee

where $\sigma=\pm 1$ has to be determined.
In order for $\psi_{\sigma}$ to be well defined, the real part of 
$b_{k,m}$ should
be positive. For each $m$, we choose $k$ such as $\Real(b_{k,m})>0$. We 
have checked numerically that there is always only one solution $k(m)$
satisfying this condition. Moreover, for $\sigma=1$ the series over
$m$ in \eref{eq11} vanishes numerically, so that 
$\sigma=-1$ gives a finite answer. 
We compute numerically $\psi_{\sigma}(b_{k,m})$ and 
$I_m(\lb)$ for every $m$, 
and plot in figure 2 the expression \rf{eq11} in the region of 
localization above $\lb=1.7$, together with data from figure 1. 
 The asymptotic curve is in agreement for $\lb > 1.9$
with the numerical results that use direct diagonalization of random matrices 
over a large number of configurations. Divergences occurs each time that
$q\lb$ is proportional to an integer. Near these points the saddle point 
approximation may not be accurate since in the series \rfb{eq11}
the terms are decreasing slowly with $m$, and we may need more correcting
terms in \eref{eq9}. 
 To get a further idea on how to simplify the series,
we see that, for large $\lb$ or large $m$, $a_{k,m}$ behaves like 

\be\label{eq14}
a_{k,m}\simeq \left (\frac{6q^2}{\lb}\right )^{1/3}\frac{1}{\phi_m}
\;\exp(2i\pi k/3).
\ee

This is a good approximation for $\lb\gg \exp(\sqrt{3/4q})\approx 1.21$ 
if $q=20$ and $m=0$.
For a ratio between the two terms inside the brackets 
in \eref{eq12} equal to 1/10, with $m=0$, we obtain a value $\lb=1.06$. 
In this approximation, $b_{0,m}=(9q\lb/2)^{1/3}$ is real and positive, and 
independent of $m$, the other solutions have negative real parts. 
We then obtain the following behavior for large $\lb$

\be\label{eq15}
I_m(\lb)\approx \sqrt{\frac{3q}{\pi}}\psi_{\sigma}
\left (\left (\frac{9q\lb}{2}
\right )^{1/3} \right )
\exp(i\sigma \pi/4)\frac{1}{\phi_m}
\ee

We find that in this limit $I_m$ is directly proportional to the inverse 
of $\phi_m$. The argument of $\phi_m$ is 

\be\label{eq16}
\arg(\phi_m)=\frac{1}{2}\arctan \left (\frac{\log \lb}{2\pi m} \right )
+\frac{\pi}{2}\theta (-m),
\ee

with $\arg(\phi_0)=\pi/4$, and $\theta(x)$ is 1 for $x>0$ 
and zero for $x\le 0$. An approximation of the probability distribution 
function is then given by

\be\nn
P(\lb\gg 1)\approx\sqrt{\frac{6}{\pi^3}}
\frac{\partial}{\partial \lb}\left \{
\frac{1}{\sqrt{|\log \lb|}}
\exp(-q\lb\log \lb+q\lb-q)\right .
\\ \label{eq17}
\left .\psi_{\sigma}\left (\left (\frac{9q\lb}{2}
\right )^{1/3} \right )
\sum_{m=-\infty}^{\infty}
\frac{\sin(2\pi q \lb m-\arg (\phi_m)+\sigma\pi/4)}{
\left (1+4\pi^2m^2/\log^2\lb\right )^{1/4}}\right \}
\ee

The term $m=0$ gives the monotonic part of the asymptotic curve,
and in order for this term not to be zero, we have to take $\sigma=-1$.
we can moreover replace $\psi_{\sigma}$ by its asymptotic value

\be
\psi_{\sigma}\left (\left (\frac{9q\lb}{2}
\right )^{1/3} \right )\approx \frac{1}{\sqrt{2}}\Gamma (3/4)
\left (\frac{2}{9q\lb}\right )^{1/4}.
\ee

A further approximation is to replace the denominator in \eref{eq17}
by $(1+\pi^2 m^2/\log^2\lb)$. The resulting function is
less accurate than the numerical integration of \eref{eq9}, but it 
should be accurate enough for very large $\lb$. Replacing the $\arctan $
\rfb{eq16} by $\pm\pi/2$, the series can be summed up using 
standard formulas. Instead, we would like to study the behavior of the 
$f(t)$ for large times, which is connected to the behavior of 
$P$ at small argument. Indeed, the main contribution of the integral in 
\eref{eq18} comes from very small $\lb$, and therefore we need the 
asymptotic behavior of $P$ in this region. 
We make the hypothesis that the saddle point equation \rfb{eq3} should 
still be valid for 
both large and small $\lb$, since the asymptotic solutions \rfb{eq4} 
in the two cases are the conjugates 
of each other using the mapping $\lb\rightarrow 1/\lb$. 
We might however modify the formula \rfb{eq17} since $\psi_{\sigma}$ has
a different behavior for small $\lb$

\be\label{eq19}
\psi_{\sigma}\left (\left (\frac{9q\lb}{2}
\right )^{1/3} \right )&\approx&
\frac{\sqrt{\pi}}{3}(1+\sigma)
\\ \nn
&-&\frac{2^{2/3}}{36}\sqrt{\pi}
\frac{\Gamma(7/12)}{\Gamma(11/12)}\frac{3+\sqrt{3}}{3-\sqrt{3}}
(\sigma-2+\sqrt{3})
\left (\frac{9q\lb}{2}
\right )^{1/3} 
\ee

and we also need to define the new arguments of the complex saddle
points

\be
\arg (\phi_m)=\frac{1}{2}\arctan \left (\frac{\log \lb}{2\pi m} \right )
-\frac{\pi}{2}\theta (-m),
\ee

with $\arg(\phi_0)=-\pi/4$. In that case, $\sigma=1$, 
so that $\psi_{1}$ is roughly constant for small $\lb$ \rfb{eq19}. 
 This result show that \rf{eq17} should still be an asymptotic solution
for small and non zero values of $\lb$, with $\psi_{1}$ instead of 
$\psi_{-1}$. This may explains the oscillations seen in the low
eigenvalue region in figure 1 (inset), for $0.2<\lb<0.5$, 
with $q=20$ and $N=800$. we can notice that, despite the fact that
the solutions from \rf{eq4} are conjugate by $\lb\rightarrow 1/\lb$,
and thus possess some symmetry,
the two regions $\lb\ll 1$ and $\lb\gg 1$ appear not to have symmetrical 
asymptotic distributions, since the global distribution itself inside
the mobility edges is not symmetric.

 A further study would be to precise the physical meaning of these 
peaks in the localized region, and this actually
appears in the structure of the real eigenvectors (see \cite{biroli.99}).
It may correspond to the particle trap in a region of low or
high connectivity (defects) as suggested by BM. An interesting model
given by BM is a Cayley tree with a defect on the central site, with a 
connectivity $c$ instead of $q+1$ for the other sites. They found some 
localized states corresponding to, for example, strong $c$. This localized 
states can disappear if a connectivity $c'$ for the surrounding neighbors 
is introduced and if $c'$ is small enough, giving rise to a {\it screening} 
effect. Also, it is not clear however why the structure is $1/q$ periodic
and if the peaks observed are Dirac peaks or are diverging with a power
law like our result suggests.
The asymptotic result \rf{eq17} is also a complement to approximations 
given in the BM's work, and also in a different way by reference \cite{dean.02}
where the central limit theorem is used to computed the fixed-point function
of an implicit equation giving the probability distribution for an effective 
Hamiltonian. 
One of the BM's approximations is based on a {\it single defect approximation}, which consists to allow fluctuations of the connectivity of a 
single site within an effective medium. They were able to find numerically the 
peaks in the localized region and gave the value of the different weights
for a given $N$ with a good approximation. In reference \cite{dean.02},
peaks seem not to diverge for large $\lb$, they appear to be rounded,
but they are observed with a shift to the right as noted by the author. 
In the small $\lb$ limit, peaks seem to be sharper, but the central limit 
theorem may not work in this case, even if the position of these peaks are 
correct. Interestingly the correct shape of the distribution in the 
delocalized region is found with high accuracy.

 I thank R\'emi Monasson for introducing me on this subject and for useful
discussions.




\begin{figure}
\label{fig1}
\begin{center}
\includegraphics[scale=0.5,angle=270]{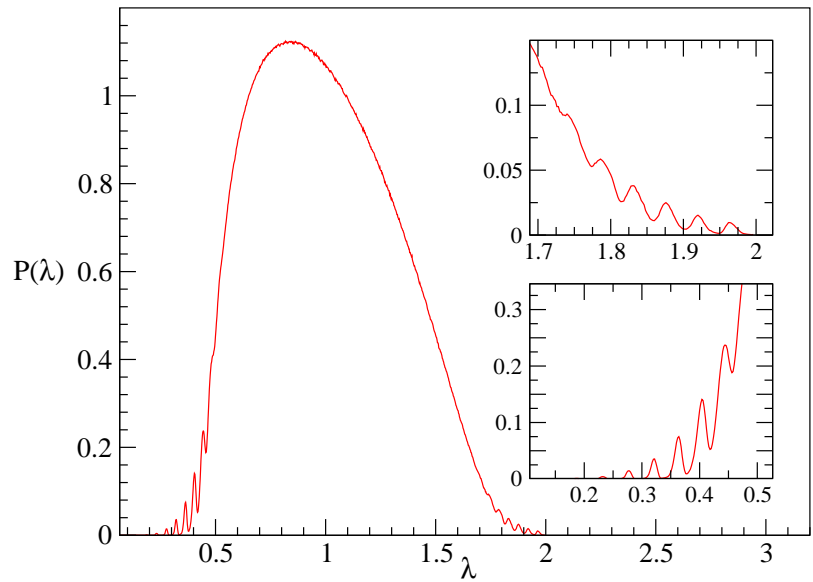}
\end{center}
\medskip
\begin{center}
\caption{Probability density function $P(\lb)$ averaged over 4500 samples
with N=800 and q=20. The inserts show oscillations at the edges of the 
distribution.}
\end{center}
\end{figure}   

\begin{figure}
\label{fig2}
\begin{center}
\includegraphics[scale=0.5,angle=270]{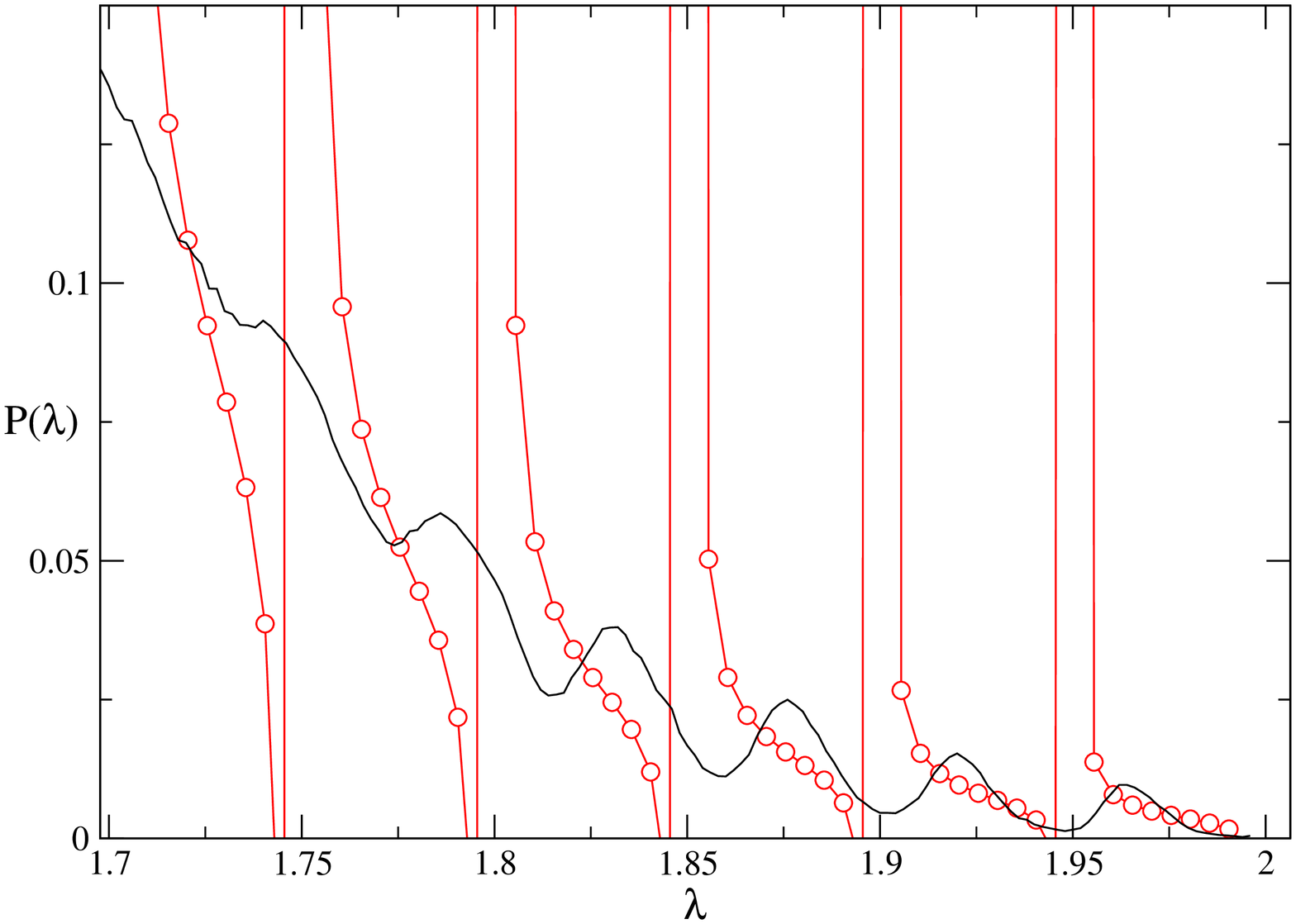}
\end{center}
\medskip
\begin{center}
\caption{Probability density function $P(\lb)$ from the asymptotic
form \rf{eq11} superposed with the numerical diagonalization curve from
figure 1 in the large $\lb$ region.}
\end{center}
\end{figure}   


\section*{References}
\begin{thebibliography}{99}

\bibitem{biroli.99}
G. Biroli and R. Monasson,
\newblock{J. Phys. A: Math. Gen.}{\bf 32}, L255 (1999).

\bibitem{campbell.85}
I.A. Campbell,
\newblock{J. Phys. Lett. (Paris)}{\bf 46}, L1159 (1985).

\bibitem{ogielski.85}
A.T. Ogielski, 
\newblock{Phys. Rev. B}{\bf 32}, 7384 (1985).

\bibitem{ogielski.87}
A.T. Ogielski, 
\newblock{Phys. Rev. B}{\bf 36}, 7315 (1987).

\bibitem{campbell.88}
I.A. Campbell, J.-M. Flesselles, R. Julien, and R. Botet,
\newblock{Phys. Rev. B}{\bf 37}, 3825 (1988).

\bibitem{svedlindh.87}
P. Svedlindh, P. Granberg, P. Nordblad, L. Lundgren, and H.S. Chen,
\newblock{Phys. Rev. B}{\bf 35}, 268 (1987).

\bibitem{bontemps.88}
N. Bontemps and R. Orbach,
\newblock{Phys. Rev. B}{\bf 37}, 4708 (1988).

\bibitem{ngai.83}
K.L. Ngai and U. Strom,
\newblock{Phys. Rev. B}{\bf 27}, 6031 (1983).

\bibitem{ngai.84}
K.L. Ngai, A.K. Rajagopal, and C.Y. Huang,
\newblock{J. Appl. Phys.}{\bf 55}, 1714 (1984).

\bibitem{ogielski.prl.85}
A.T. Ogielski and D.L. Stein,
\newblock{Phys. Rev. Lett.}{\bf 55}, 1634 (1985).

\bibitem{liu.92}
C.H. Liu and S.R. Nagel,
\newblock{Phys. Rev. Lett.}{\bf 68}, 2301 (1992).

\bibitem{jia.99}
X. Jia, C. Caroli, and B. Velicky,
\newblock{Phys. Rev. Lett.}{\bf 82}, 1863 (1999).

\bibitem{leibig.94}
M. Leibig,
\newblock{Phys. Rev. E}{\bf 49}, 1647 (1994).

\bibitem{viana.85}
L. Viana and A.J. Bray,
\newblock{J. Phys. C: Solid State Phys.}{\bf 18}, 3037 (1985).

\bibitem{bray.88}
A.J. Bray and G.J. Rodgers,
\newblock{Phys. Rev. B}{\bf 38}, 11 461 (1988).

\bibitem{kim.85}
Y. Kim and A.B. Harris,
\newblock{Phys. Rev. B}{\bf 31}, 7393 (1985).

\bibitem{rodgers.88}
G.J. Rodgers and A.J. Bray,
\newblock{Phys. Rev. B}{\bf 37}, 3557 (1988).

\bibitem{kanter.87}
I. Kanter and H. Sompolinsky,
\newblock{Phys. Rev. Lett.}{\bf 58}, 164 (1987).

\bibitem{dean.02}
D.S. Dean
\newblock{J. Phys. A:Math. Gen.}{\bf 35}, L153 (2002)




\end{thebibliography}
\end{document}